%% file: main.tex
\documentclass[a4paper,11pt]{article}
\usepackage{jheppub} 
\usepackage{lineno}
\usepackage{slashed}
\usepackage{booktabs}
\usepackage{arydshln}
\usepackage{amssymb}
\usepackage{amsthm}
\usepackage{cancel}
\usepackage{slashed}
\usepackage{bm}
\usepackage{dsfont}
\usepackage{amsmath}
\usepackage{booktabs}
\usepackage{multirow}
\usepackage{nicematrix}
\usepackage{siunitx}
\DeclareMathOperator{\Br}{Br}

\usepackage{xcolor}
\usepackage{color}
\usepackage{subcaption}
\usepackage{enumitem}

\definecolor{myblueold}{RGB}{65,105,225}
\definecolor{myblue}{RGB}{0, 71, 171}
\definecolor{mahogany}{RGB}{156, 0, 0}
\definecolor{mygray}{RGB}{145,145,145}
\definecolor{mygreen}{RGB}{34,139,34}
\definecolor{mybrown}{RGB}{110, 38, 14}
\definecolor{mypurple}{RGB}{218,112,214}

\def\fn{\footnote}
\newcommand{\ali}[1]{\begin{align}#1\end{align}}

\newcounter{GHQ}

\newcounter{JBQ}

\newcounter{DFQ}

\allowdisplaybreaks

\title{\boldmath Do large QCD corrections to di-Higgs decay survive parton showering? A study of $HH\to b\bar{b}\gamma\gamma$
}

\author[a]{Jens Braun,}
\author[a]{Duarte Fontes,}
\author[a]{Gudrun Heinrich}
\affiliation[a]{Institute for Theoretical Physics,
Karlsruhe Institute of Technology, Wolfgang-Gaede-Str.1, \\
76131 Karlsruhe, Germany}

\date{} 

\emailAdd{j.braun@kit.edu}
\emailAdd{duarte.fontes@kit.edu}
\emailAdd{gudrun.heinrich@kit.edu}

\abstract{
While significant effort has been devoted to precision calculations of the production of two Higgs bosons via gluon fusion, the treatment of their decays in this process has only recently begun to attract attention. It has been found that fixed-order QCD corrections to fiducial di-Higgs decay rates involving the $b\bar{b}$ decay channel can be substantial. Considering $HH\to b\bar{b}\gamma\gamma$, we show that such corrections arise predominantly from sensitivity to soft and collinear QCD radiation at fixed order, and that they are largely washed out once parton showers are included.
}

\makeatletter
\gdef\@fpheader{}
\makeatother

\begin{document}

\begin{flushright}
KA-TP-24-2025,\\
P3H-25-61
\end{flushright}
\vspace*{1cm}

\maketitle
\flushbottom

\section{Introduction}

\input{intro.tex}

\section{Framework}
\label{sec:framework}

\input{framework.tex}

\section{Results}
\label{sec:results}

\input{results.tex}

\section{Conclusions}
\label{sec:conclusions}

\input{conclusions.tex}

\section*{Acknowledgments}
\input{acknowledgments.tex}




\bibliographystyle{JHEP}
\bibliography{refs.bib}

\end{document}

%% file: intro.tex
Higgs boson pair production offers exceptional promise for uncovering new physics. Indeed, if the Higgs trilinear coupling deviates from the Standard Model (SM) prediction, di-Higgs production could reveal hints of physics beyond the SM before a new particle is detected. On the other hand, such deviations can only be clearly identified as signposts of new physics if the SM prediction for this process are under very good theoretical control.

In gluon-fusion di-Higgs production, the main sources of theoretical uncertainty arise from missing higher-order QCD and electroweak (EW) corrections, approximations such as the $m_t \to \infty$ limit, and scheme-dependent treatments of the top-quark mass. Considerable effort has already been devoted to reducing these uncertainties. The leading order (LO) cross section has been calculated in refs.~\cite{Glover:1987nx,Eboli:1987dy,Plehn:1996wb}; next-to-leading order (NLO) QCD corrections in the heavy top limit (HTL) have been presented in ref.~\cite{Dawson:1998py};
NLO QCD corrections including the full top-quark mass dependence have been calculated in refs.~\cite{Borowka:2016ehy,Borowka:2016ypz,Baglio:2018lrj,Baglio:2020ini}. 
The results of refs.~\cite{Borowka:2016ehy,Borowka:2016ypz} have been matched to parton showers in refs.~\cite{Heinrich:2017kxx,Heinrich:2019bkc}, and combined with next-to-next-to-leading (NNLO) corrections in the so-called $\mathrm{NNLO_{FTapprox}}$ framework, where only the virtual contributions are evaluated in the HTL~\cite{Grazzini:2018bsd}. These results have also been incorporated into calculations at N$^3$LO in the HTL~\cite{Chen:2019lzz,Chen:2019fhs} and in combined N$^3$LO+N$^3$LL predictions~\cite{AH:2022elh}.
Analytic approaches for the NLO QCD corrections to Higgs boson pair production, based on increasingly accurate approximations that cover nearly the entire phase space~\cite{Bellafronte:2022jmo,Davies:2023vmj}, have also become available~\cite{Davies:2018qvx,Davies:2019dfy,Bagnaschi:2023rbx,Davies:2025qjr}.

In recent years, therefore, significant progress has been made in reducing both scale uncertainties and those arising from approximations of the top-quark mass dependence. 
Uncertainties from EW corrections have also been substantially reduced following the computation of the full NLO EW corrections in ref.~\cite{Bi:2023bnq}. These corrections have additionally been obtained in a $1/m_t$ expansion~\cite{Davies:2023npk}; see also refs.~\cite{Borowka:2018pxx, Bizon:2018syu, Muhlleitner:2022ijf,Davies:2022ram,Bizon:2024juq,Li:2024iio} for earlier partial EW corrections. Analytic results are available for the light-quark contributions~\cite{Bonetti:2025vfd}, as well as for the Yukawa and Higgs-self-coupling corrections in the high-energy limit~\cite{Davies:2025wke}. Numerical results for the Yukawa and self-coupling contributions have been presented in ref.~\cite{Heinrich:2024dnz}.
Currently, the uncertainties that dominate the overall uncertainty budget are those associated with the choice of top-quark mass renormalization scheme --- first identified in refs.~\cite{Baglio:2018lrj,Baglio:2020ini} and further studied in refs.~\cite{Jaskiewicz:2024xkd,Davies:2025qjr}. This situation is expected to improve in the near future. In fact, the large logarithms responsible for the difference between the on-shell and $\overline{\text{MS}}$ schemes at high energies have been identified in ref.~\cite{Jaskiewicz:2024xkd}, and ongoing efforts to compute NNLO corrections with full top-quark mass dependence~\cite{Davies:2023obx,Davies:2024znp,Davies:2025ghl} are likely to further reduce these uncertainties.

All this remarkable progress has been directed towards reducing the theoretical uncertainties of gluon-fusion di-Higgs \textit{production}. This naturally raises the question of the impact of the subsequent Higgs boson \textit{decays}. In other words, with production now brought under substantially improved theoretical control, do corrections to the Higgs boson decays compromise this accuracy by introducing sizable effects?
For a long time, this question remained unaddressed in the context of Higgs boson pair production. This situation changed with ref.~\cite{Li:2024ujf}, which investigated gluon-fusion di-Higgs production followed by decay into $b \bar{b} \gamma\gamma$ within the narrow-width approximation (NWA). The study found that NLO QCD corrections to the decay $H \to b\bar{b}$ can induce large effects, of order 19\% for the fiducial integrated cross section. In a later work, the same authors considered the final state $b \bar{b} \tau^+ \tau^-$ and reached similar conclusions \cite{Li:2025gbx}.

Here, we investigate the origin of these large corrections. We show that they are an artifact of fixed-order (FO) calculations, where infrared (IR) sensitivity leads to large logarithms. Parton showers (PS), by resumming these logarithms, mitigate the effect: the large NLO QCD corrections to di-Higgs decays observed at FO are therefore washed out. Similar observations were recently reported in ref.~\cite{Behring:2025msh}, which studied Higgs boson decays to $b\bar{b}$ in weak boson fusion, comparing FO NNLO results~\cite{Asteriadis:2021gpd, Asteriadis:2024nbg} with MiN(N)LO predictions~\cite{Hamilton:2012rf,Bizon:2019tfo}.%
\fn{Studies of $H\to b\bar{b}$ at higher orders in QCD in the context of associated production have been carried out in refs.~\cite{Gauld:2019yng,Ferrera:2017zex,Caola:2017xuq,Campbell:2016jau}.}
Ref.~\cite{Behring:2025msh} concluded that the large FO corrections are reduced once PS are included. In the present study, we illustrate this phenomenon in the channel $gg \to HH \to b \bar{b}\,\gamma\gamma$, which combines a clean diphoton signal with the large branching fraction of $H \to b\bar b$, and which has been the subject of dedicated searches by both ATLAS~\cite{ATLAS:2025hhd} and CMS~\cite{CMS:2020tkr}. Furthermore, we show that the same phenomenon persists when the $\gamma\gamma$ system is replaced by any other final state involving no colored particles.

The remainder of the paper is structured as follows. In section~\ref{sec:framework}, we outline the framework of our calculation, explaining the factorization of the cross section and its implementation in our codes. Section~\ref{sec:results} presents our results and section~\ref{sec:conclusions} summarizes our findings.

%% file: framework.tex
We investigate NLO QCD corrections to the di-Higgs decays in $gg \to HH \to b \bar{b} \gamma \gamma$, as well as the effects of the inclusion of PS in the process. We start by discussing preliminary theoretical aspects of this investigation in section \ref{sec:theory}, and we then turn to its implementation in section \ref{sec:implementation}.

\subsection{Theory considerations}
\label{sec:theory}

 We work in the NWA, which implies a factorization of production and decay. The Higgs bosons are thus assumed to be on-shell, and the QCD corrections for production and decay are completely separable. The differential cross section for gluon fusion di-Higgs production and decay to $b \bar{b} \gamma \gamma$ can then be written as
\ali{
\label{eq:NWA-specific}
d\sigma
= d\sigma_{\text{ggF}}\;
2 \, 
\mathrm{Br}(H\to b \bar{b})\, d\gamma_{b \bar{b}}\;
\mathrm{Br}(H\to \gamma\gamma)\, d\gamma_{\gamma\gamma}.
}
$d\sigma_{\text{ggF}}$ is the differential di-Higgs production cross section via gluon fusion, 
the factor of 2 accounts for the fact that both Higgs bosons can decay into $b \bar{b}$ and into $\gamma \gamma$, 
$\mathrm{Br}(H\to X)$ is the $H \to X$ branching ratio, and
\ali{
\label{eq:dgamma-def}
d\gamma_{X} \;\equiv\; \frac{d\Gamma(H\to X)}{\Gamma(H\to X)}.
}
Here $\Gamma(H\to X)$ is the partial width for the channel $X$ and $d\Gamma(H\to X)$ the corresponding Lorentz-invariant differential decay matrix element. If $X$ consists solely of colorless particles, no real gluon emission arises at NLO, a direct consequence of color conservation. In particular, there is no NLO QCD radiation in $H \to \gamma\gamma$, so that all NLO QCD corrections to $d\gamma_{\gamma\gamma}$ are virtual.
This implies the factorization of $HH \to b \bar{b} \gamma \gamma$ into $H \to b \bar{b}$ and $H \to \gamma\gamma$ implicit in eq.~(\ref{eq:NWA-specific}).
It also implies that the (virtual) NLO QCD corrections to $d\gamma_{\gamma\gamma}$ cancel in the normalized ratio $d\gamma_{\gamma\gamma} \equiv d\Gamma(H\to\gamma\gamma)/\Gamma(H\to\gamma\gamma)$; indeed, if fiducial selections $F$ are applied to the photons, we find
\ali{
\label{eq:gammagamma}
\int d\gamma_{\gamma\gamma}
=\frac{\int F\, d\Gamma(H\to\gamma\gamma)}{\Gamma(H\to\gamma\gamma)}
= \frac{|\mathcal{M}_{H\to\gamma\gamma}|^2}{|\mathcal{M}_{H\to\gamma\gamma}|^2} \frac{\int F d \Phi}{\int d \Phi} = \frac{\Omega^{\text{fid}}}{\Omega},
}
with $\Omega^{\text{fid}}$ and $\Omega$ being the phase space volume with and without fiducial cuts, respectively. Therefore, the integral in eq.~(\ref{eq:gammagamma}) is purely an acceptance (unity if $F\!=\!1$) and receives no QCD corrections at NLO.

Two important conclusions follow from these observations. The first is that NLO QCD effects in $HH \to b \bar{b} \gamma \gamma$ arise solely from $H\to b\bar b$. This implies, in the second place, that $\gamma\gamma$ may be replaced by another final state $X$ made exclusively of colorless particles (e.g.\ $\tau^+\tau^-$, $ZZ^{(*)}\!\to4\ell$, $WW^{(*)}\!\to\ell\nu\ell\nu$, $Z\gamma$): the corresponding $d\gamma_{X}$ has no QCD correction at NLO, and its fiducial integral is simply a ratio of phase space volumes. As a consequence, the conclusions derived by ref.~\cite{Li:2024ujf} about large NLO QCD corrections extend straightforwardly to the case $HH \to b\bar{b} \tau^{-} \tau^{+}$ discussed by the same authors in ref.~\cite{Li:2025gbx}.

Finally, we emphasize that our focus lies on the NLO QCD corrections to the di-Higgs \textit{decays} in the process $gg \to HH \to b \bar{b} \gamma \gamma$. Our goal is not to provide a complete NLO QCD calculation of the process, but rather to investigate the origin of the large corrections reported in refs.~\cite{Li:2024ujf,Li:2025gbx}, and to determine whether these corrections related to the decay persist once PS effects are included. The impact of PS in combination with NLO QCD corrections to di-Higgs \textit{production} has already been investigated in the literature~\cite{Heinrich:2017kxx,Heinrich:2019bkc}. Furthermore, within the NWA at NLO QCD, including NLO QCD corrections to the production would require treating the di-Higgs decays at LO, which would defeat the purpose of our study. To avoid introducing unnecessary complications, we therefore restrict our analysis to QCD corrections in the decays.

\subsection{Implementation}
\label{sec:implementation}

\begin{figure}[h!]
\centering
\includegraphics[width=1\textwidth]{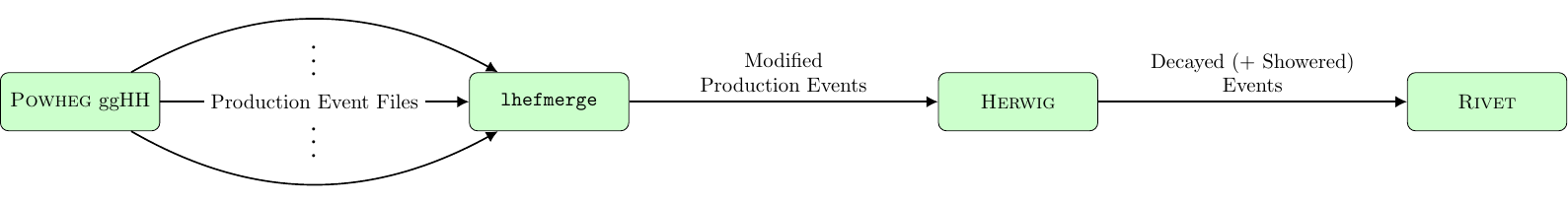}
\caption{Workflow of the event generation and analysis setup.}
\label{fig:Chart}
\end{figure}

We generated several sets of Monte Carlo samples for $gg \to HH \to b\bar{b} \gamma\gamma$.
We considered both LO and NLO descriptions of $H \to b \bar{b}$, each evaluated either at FO or matched to PS. Figure \ref{fig:Chart} illustrates the workflow through which we generated the different samples. In what follows, we explain in detail the different elements.

Starting with production, we consider proton-proton collisions at the center-of-mass energy $\sqrt{s}=14$~TeV. Di-Higgs production through gluon fusion is simulated with \textsc{Powheg} \cite{Nason:2004rx,Frixione:2007vw,Alioli:2010xd}, retaining the full top-mass dependence in the loop amplitudes~\cite{Borowka:2016ehy,Borowka:2016ypz,Heinrich:2017kxx}. As discussed above, we restrict our investigation of the production to the LO results in $\alpha_s$. 
Moreover, we consider the $b$ quark to be massless in the production; at the level of the total cross section, this is a reasonable approximation, the difference being below the percent level \cite{Maltoni:2014eza}.
Parton distribution functions (PDFs) and strong coupling constant are taken from \verb|PDF4LHC15_nlo_100_pdfas|~\cite{Butterworth:2015oua}, and their evolution is obtained from LHAPDF \cite{Buckley:2014ana}. Following ref.~\cite{Li:2024ujf}, the (production) renormalization scale $\mu_{R,p}$ and factorization scale $\mu_F$ are chosen to be equal, $\mu_{R,p} = \mu_F =  m_{HH}/2$, with $m_{HH}$ being the di-Higgs invariant mass. The dependence of the results on $\mu_{R,p}$ and $\mu_F$ is well known~\cite{Heinrich:2017kxx} and, given our focus on the decays, is not investigated in what follows.

Once the output from \textsc{Powheg} is obtained, we process the generated events with a custom routine, \texttt{lhefmerge}, which serves two purposes: first, it merges the event files from parallelized \textsc{Powheg} runs into a single file; second, it modifies the identifier of the second Higgs boson in each event, replacing it with a fictitious particle $H_1$. This particle is defined to be identical to the Higgs boson $H$, except that it is constrained to decay exclusively as $H_1 \to \gamma\gamma$, while the original Higgs boson is forced to decay as $H \to b\bar{b}$. This modification is introduced solely to enhance efficiency, since otherwise only a very small fraction of events would result in the final state $HH \to b\bar{b}\gamma\gamma$. The physical results remain unaffected, as they depend only on the fraction of $HH \to b\bar{b}\gamma\gamma$ events passing the fiducial cuts, not on their total number.

Higgs boson decays and the PS are performed with \textsc{Herwig}  \cite{Bellm:2015jjp,Bellm:2017bvx,Bellm:2019zci,Bewick:2023tfi}, with hadronization and the underlying event disabled. All necessary frame transformations are performed internally by \textsc{Herwig}, which provides built-in implementations of the LO decay $H \to \gamma\gamma$ and the NLO decay $H \to b\bar b$.
$H \to \gamma\gamma$ is trivially implemented, as discussed in the previous section. 
For $H\to b\bar b$ we consider both the LO case, as well as the case with NLO QCD corrections, the latter evaluated entirely in the on-shell scheme.
The only explicit scale dependence therefore shows up in renormalization scale $\mu_{R,d}$ that controls the evaluation of $\alpha_s(\mu_{R,d})$, which we set to $\mu_{R,d}=m_H$. 
We note that ref.~\cite{Li:2024ujf} instead adopts $\mu_{R,d}=\mu_{R,p}=m_{HH}/2$. Although this choice is technically consistent with the NWA, we consider $\mu_{R,d}=m_H$ to be the more natural scale for the decay, as it reflects the characteristic hard scale of the $H \to b\bar{b}$ process. A discussion of the estimation of theoretical uncertainties in the decays is deferred to section~\ref{sec:results}.

Still concerning $H\to b\bar b$, we assume a massive $b$ quark in the matrix elements and in the phase space, with mass $m_b(m_b) = 4.18$ GeV.
Here again we differ from ref.~\cite{Li:2024ujf}, which considers a massless-$b$ setup.
A finite $m_b$ is not only employed in the \textsc{Herwig} implementation of $H \to b\bar b$ at NLO, but also ensures a stable interface to the shower. The differences between massive and massless $b$ should be of subpercent level \cite{Asteriadis:2024nbg}.
We note that the Yukawa coupling ends up cancelling in the ratio $d\gamma_{b\bar{b}}$.
IR singularities in the FO calculation are handled with the default Catani--Seymour dipole subtraction \cite{Catani:1996jh,Catani:1996vz} in \textsc{Matchbox} \cite{Platzer:2011bc}. When we turn on PS, showering is performed with the angular-ordered $\tilde q$ algorithm~\cite{Gieseke:2003rz}.

Analyses are performed with \textsc{Rivet} \cite{Bierlich:2019rhm}, which calls \textsc{FastJet} \cite{Cacciari:2011ma} for jet clustering, where we use the anti-$k_T$ algorithm~\cite{Cacciari:2008gp}. Unless stated otherwise, we set the jet radius to $R=0.4$.
We use the default \textsc{Rivet} prescription for parton-level $b$-tagging. 
To unambiguously identify the two $b$-jets from the Higgs boson decay, we select events with exactly two $b$-jets, defined by $p_T^j \geq 25~\text{GeV}$ and $|\eta^j| \leq 2.5$ (we place no constraints on additional jets). The invariant mass of the two $b$-jets is further required to satisfy $90~\text{GeV} \le m_{jj} \le 190~\text{GeV}$. For the photons, we impose $p_T^{\gamma} \geq 25~\text{GeV}$ and $|\eta^{\gamma}| \leq 2.5$, together with separation cuts $\Delta R_{jj}, \Delta R_{j\gamma}, \Delta R_{\gamma\gamma} \geq 0.4$, in order to ensure well-isolated $b$-jets and photons, and avoid overlapping objects in the fiducial phase space.%
\fn{$\Delta R_{j\gamma}$ is required for each of the four possible pairings between one of the two $b$-jets and one of the two photons.}
The selection described in this paragraph is intended to reproduce, as closely as possible, the strategy of ref.~\cite{Li:2024ujf}, and is broadly consistent with the CMS strategy in the $HH \to b\bar b\gamma\gamma$ channel~\cite{CMS:2020tkr}.

The fiducial cross section is
\ali{
\sigma = 2 \, \mathrm{Br}(H\!\to\! b \bar b) \, \mathrm{Br}(H\!\to\!\gamma\gamma) \, \sigma_{\text{ggF}} \, Q,}
with $\sigma_{\text{ggF}}$ being the total production cross section via gluon fusion, and
\ali{
Q \equiv \frac{1}{\sigma_{\text{ggF}}}\int 
F(\Phi_{HH};\gamma_{b\bar b},\gamma_{\gamma\gamma})\,
d\sigma_{\text{ggF}}(\Phi_{HH})\,d\gamma_{b \bar b}\,d\gamma_{\gamma\gamma}.
}
$F$ encodes the fiducial cuts; $\Phi_{HH}$ denotes the Higgs-pair production phase space, while $\gamma_{b\bar b}$ and $\gamma_{\gamma\gamma}$ parameterize the decay kinematics in the Higgs rest frames. In practice, we evaluate $Q$ as the acceptance, i.e.
\ali{ \label{eq:Q}
Q \;=\; \frac{N_{\text{pass}}}{N_{\text{total}}},
}
where $N_{\text{total}}$ is the total number of generated events and $N_{\text{pass}}$ the number of events that satisfy all fiducial cuts.
We set the numerical values of the SM input parameters to
\ali{
\label{eq:inputs}
m_H &= \qty{125}{\GeV},
&
m_t &= \qty{173}{\GeV}, \nonumber \\
\Br\left(H \to b\bar{b}\right) &= \num{0.5824},
&
\Br\left(H \to \gamma\gamma\right) &= \num{2.27e-3},
}
in agreement with ref.~\cite{Li:2024ujf}.%
\fn{While in principle the branching ratios could have corrections, we set them to the fixed values given in eq.~(\ref{eq:inputs}).}

%% file: results.tex
We finally turn to our numerical results. As a reference, the (LO) production cross section amounts to
$
\sigma_\mathrm{ggF} = \num{19.873 +- 0.002} \, \mathrm{fb},
$
where the quoted uncertainty reflects the Monte Carlo integration. 
While this value simply provides the normalization for our study, our main interest lies in the impact of QCD corrections to the Higgs boson decays. 
To this end, table~\ref{tab:xsec-table} reports the fiducial cross sections obtained with the decay treated at LO and at NLO, both without PS (i.e.~at FO) and including the showers. Again, the displayed uncertainty is the Monte Carlo uncertainty, calculated by combining the propagated uncertainty from the production cross section with Poisson uncertainties on the number of counted events used to calculate $Q$ from eq. \eqref{eq:Q}.
The fiducial selection is the one defined in the previous section. 
For each case, we also display the factor $K \equiv \sigma_{\text{NLO-decay}}/\sigma_{\text{LO-decay}}$,
which directly quantifies the size of the NLO corrections in the decay.
\begin{table}[h!]
  \centering
  \bigskip
  \begin{tabular}{l S S}
    \toprule
     & {FO} & {FO+PS} \\
    \midrule
    $\sigma_{\text{LO-decay}}$ (fb) & \num{0.02783 +- 0.00005}  & \num{0.02358 +- 0.00004} \\
    $\sigma_{\text{NLO-decay}}$ (fb) & \num{0.02363 +- 0.00004} & \num{0.02289 +- 0.00004} \\\midrule
    \multicolumn{1}{c}{$K$} & \num{0.849 +- 0.002} & \num{0.971 +- 0.003}\\
    \bottomrule
  \end{tabular}
  \caption{Fiducial cross section for $gg \to HH \to b \bar{b} \gamma \gamma$, assuming the selection described in section \ref{sec:implementation}. See text for details.}
  \label{tab:xsec-table}
\end{table}

Before discussing the results, we comment on the lack of uncertainties beyond the Monte Carlo ones, both for the FO results and for the results with showers.
Concerning the former, it is customary to estimate the theoretical uncertainty by varying the renormalization scale up and down by a factor of two. In our case, and as mentioned above, the only explicit scale dependence arises through $\alpha_s(\mu_{R,d})$. Varying $\mu_{R,d}$ around our central choice $m_H$ (taking $\mu_{R,d}=2m_H$ and $\mu_{R,d}=m_H/2$), we observe differences of order $\mathcal{O}(1\%)$. Two remarks follow: first, this variation is much smaller than the uncertainties affecting the production process \cite{Li:2024ujf}; second, the FO result is not particularly meaningful in itself, since (as we shall see) it is affected by large logarithms that are resummed once PS effects are included. In this sense, performing scale variations on a result that is intrinsically unphysical is of limited relevance. As for the results with PS effects, estimating the theoretical uncertainty is a more involved task. While such an analysis is certainly interesting in its own right, it is not the focus of the present work, which is instead to demonstrate that the large FO corrections are largely washed out by parton showers. For these reasons, we do not pursue a systematic uncertainty estimate in what follows.

Focusing now on the results of the table, the $K$ factor at FO is $0.849$, corresponding to a relative correction of $-15.1\%$ with respect to LO. This is smaller in magnitude than the correction of about $-19\%$ reported in ref.~\cite{Li:2024ujf}. We attribute this discrepancy to differences in the renormalization scale, renormalization scheme and mass definitions adopted in the two calculations.
For the purposes of our analysis, the precise numerical value is of secondary importance; the essential point is that in both cases the corrections are sizable. 
In this sense, we qualitatively confirm the conclusion of ref.~\cite{Li:2024ujf}, namely that NLO QCD corrections to the di-Higgs decays in $gg \to HH \to b \bar{b} \gamma \gamma$ induce large effects on the fiducial cross section. 

Table~\ref{tab:xsec-table} also shows that the sizable negative correction observed at FO of $-15.1\%$ is drastically reduced once PS are included. 
With showering, indeed, the corresponding correction is only $-2.9\%$, i.e.~almost an order of magnitude smaller. 
This demonstrates that the large FO effects in the fiducial cross section, arising from the NLO treatment of $H \to b\bar b$, are to a large extent washed out by the inclusion of PS. 
This constitutes the central finding of our study.

\begin{figure}[h!]
    \centering
    \begin{subfigure}[l]{0.49\textwidth}
        \includegraphics[width=\textwidth]{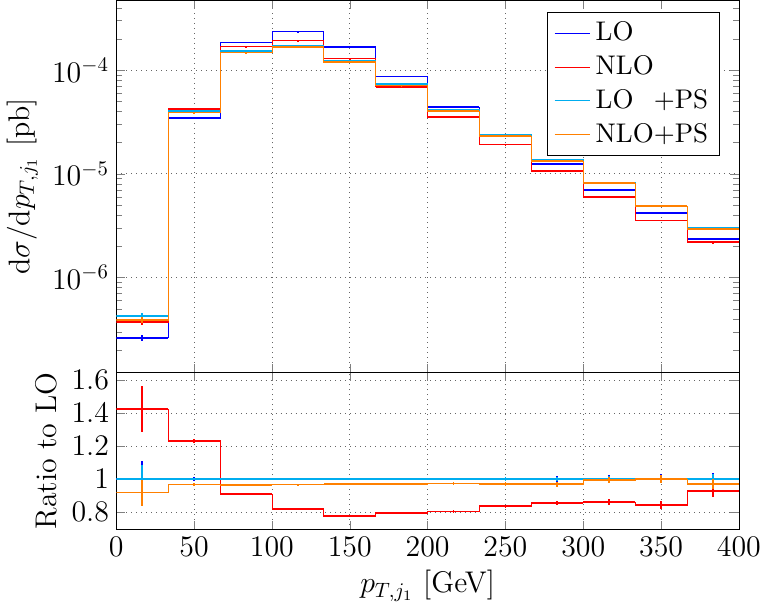}
    \end{subfigure}
    \begin{subfigure}[r]{0.49\textwidth}
        \includegraphics[width=\textwidth]{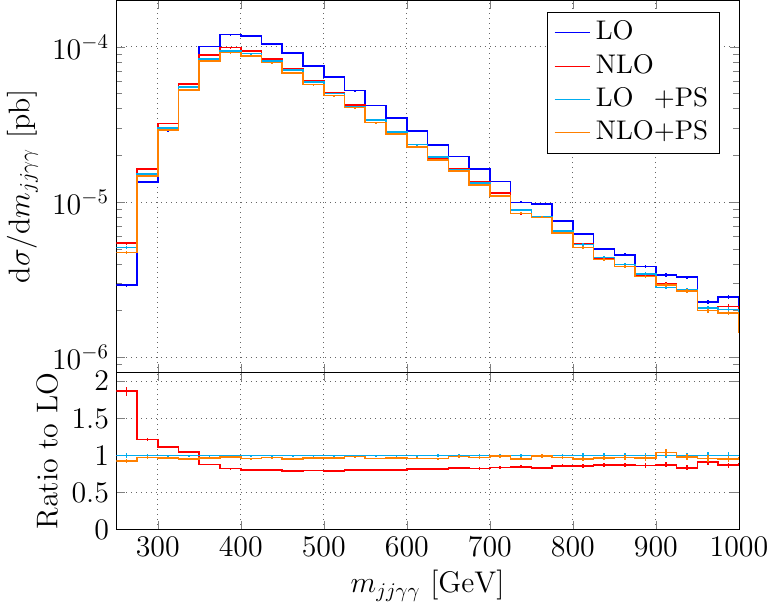}
    \end{subfigure}
    \caption{Distributions of the transverse momentum of the leading jet (left) and the invariant mass of the two $b$-jets and two photons (right), assuming the fiducial selection described in section \ref{sec:implementation}. In each plot, the upper pane shows results with the production evaluated at LO; the $H \to b \bar{b}$ decay is calculated either at LO or at NLO, and with or without PS. The lower pane of each plot shows the ratio of the different curves of the upper pane to the corresponding LO curve.}
    \label{fig:distributions}
\end{figure}

These conclusions are not restricted to the integrated fiducial cross section of table~\ref{tab:xsec-table}, but also hold at the level of differential distributions. To illustrate this, we show in figure~\ref{fig:distributions} the cross sections differential in the transverse momentum of the leading $b$-jet (left panel) and in the invariant mass of the two $b$-jets and two photons, $m_{jj\gamma\gamma}$ (right panel).
In both observables, the pattern is the same as for the integrated result: at FO, the NLO corrections are for the most part sizable with respect to LO, while once the PS are included, the LO and NLO predictions become nearly indistinguishable. 

To better understand the origin of the large FO NLO corrections in the decays, we now examine their dependence on two key fiducial parameters: the transverse-momentum cut $p_T^{\min}$ applied to the $b$-jets and the jet radius $R$ used in the clustering algorithm. Both parameters crucially affect fiducial cross sections, as they control the acceptance of real radiation. It is therefore natural to expect FO predictions to vary strongly with their values, and indeed we will see that the NLO corrections change substantially when these parameters are pushed to extreme regions of phase space. In what follows, we outline the expected behavior for each variable.  

For the jet radius $R$, it has long been known that cross sections allowing more than one parton per jet develop a dependence on $\ln R$~\cite{Ellis:1992qq}. Similar arguments hold for photon isolation cones~\cite{Catani:2013oma}, where resummation of large logarithms of the cone size is required (see e.g.~ref.~\cite{Fontannaz:2025dps}).
Intuitively, for very small $R$, NLO radiation often falls outside the jet cone, so the $b$-jets lose transverse momentum relative to LO. This out-of-cone loss enhances the sensitivity to the jet definition and leads to sizable corrections. As $R$ increases, more of the final-state radiation is captured inside the jets, reducing the relative size of the corrections and stabilizing the predictions.  

A similar reasoning applies to the minimal transverse momentum cut on the $b$-jets, $p_T^{\min}$. At LO, in the Higgs boson rest frame, the $b$-quarks from $H\to b\bar b$ carry a fixed momentum of $m_H/2$. Since in $gg\to HH$ the Higgs bosons are only moderately boosted, the $b$-jet transverse momenta in the lab frame remain centered around this scale, with limited broadening. If $p_T^{\min}$ is set well below this characteristic value, essentially all $b$-jets pass the selection already at LO, so NLO corrections have little impact. In contrast, if $p_T^{\min}$ is chosen above this scale, only a small fraction of LO events survive. At NLO, additional gluon emission smears the $b$-quark energies and can easily push the $b$-jet $p_T$ below $p_T^{\min}$, producing large relative corrections.

\begin{figure}[h!]
    \centering
    \begin{subfigure}[l]{0.49\textwidth}
        \centering
        \includegraphics[width=\textwidth]{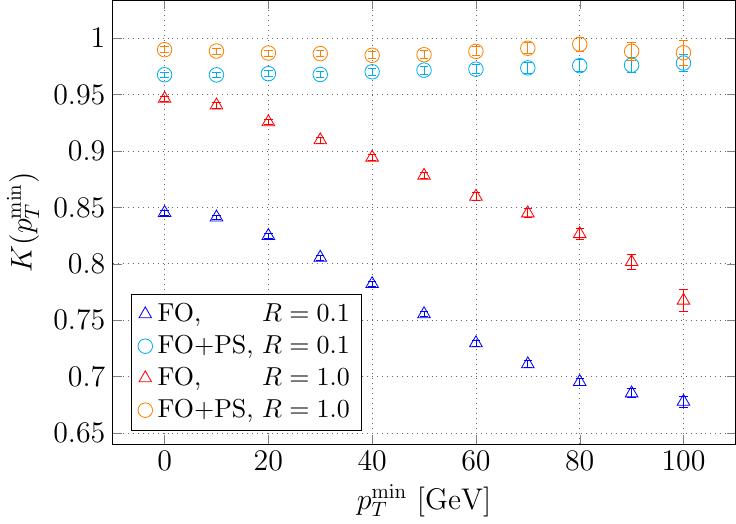}
    \end{subfigure}
    \begin{subfigure}[r]{0.49\textwidth}
        \centering
        \includegraphics[width=\textwidth]{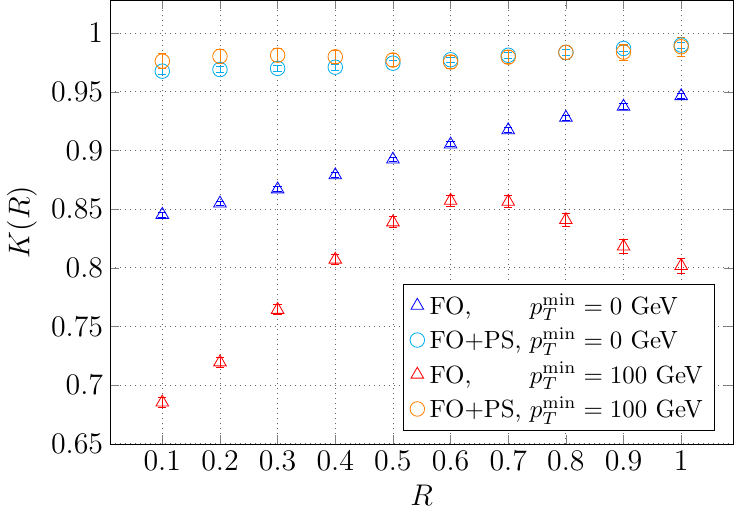}
    \end{subfigure}
    \caption{The variation of $K$ (ratio of fiducial cross sections with decays at NLO and at LO), as a function of the $p_T^\mathrm{min}$ cut (left) and of the jet radius parameter $R$ (right). The vertical bars represent uncertainties due to Monte Carlo integration.}
    \label{fig:ptmin_R_scans}
\end{figure}

These expectations are clearly confirmed by the behavior shown in figure~\ref{fig:ptmin_R_scans}. Both panels display the relative difference between fiducial cross sections evaluated with NLO decays and with LO decays, expressed through the ratio $K$.
In all cases, the NLO corrections are negative, reflecting the fact that additional gluon emission tends to lower the reconstructed $b$-jet transverse momenta, thereby reducing the number of events that pass the fiducial selection.  

The left panel illustrates this dependence very clearly. For $R=1$ (red triangles) and very low $p_T^{\min}$, the FO corrections are small, since virtually all jets pass the selection both at LO and NLO. As $p_T^{\min}$ grows, the corrections grow in magnitude (more negative), consistent with the expectation that even modest recoil from an emitted gluon can push a $b$-jet below the cut. The comparison between the two radii underscores the role of out-of-cone losses: for the smaller radius ($R=0.1$, blue triangles) the NLO corrections are markedly more negative than for the larger radius, because radiation is more likely to escape the narrow jet cone.

The right panel reorganizes the same information as a function of $R$ at fixed $p_T^{\min}$. A similar pattern is observed: small radii yield stronger negative corrections following the $\ln{R}$ behavior, while larger radii capture more radiation and therefore reduce the size of the corrections. This monotonic behaviour holds for $p_T^{\min}=0$~GeV (blue triangles). By contrast, for $p_T^{\min}=100$~GeV (red triangles) two regimes emerge: for $R\lesssim 0.6$ the recovery of emitted radiation dominates and $K$ rises with increasing $R$, whereas for $R\gtrsim 0.6$ $K$ decreases. In this large-$R$ regime, finite-angle gluons radiated off one $b$ are increasingly clustered into the \emph{other} $b$-jet. This depletes the emitter-jet $p_T$ and
drives NLO-decay events below $p_T^{\min}$, while LO-decay events have no analogous loss. Consequently, the NLO acceptance drops with $R$ and $K(R)$ decreases for large radii.

Finally, in both panels the predictions matched to PS (circles) exhibit two salient features: (i) the NLO-to-LO shifts are small and (ii) the results are essentially stable under variations of both $R$ and $p_T^{\min}$, remaining virtually flat across the scans. We also note that the Monte Carlo uncertainties (vertical bars) increase with $p_T^{\min}$, reflecting the progressively lower selection efficiency discussed above: as the cut hardens, fewer events pass, and the corresponding statistical errors grow.

%% file: conclusions.tex
After substantial theoretical progress in gluon fusion di-Higgs production, the associated perturbative uncertainties in the production cross section have been substantially reduced. More recently, it was pointed out that fixed-order (FO) NLO QCD corrections to the decays can induce large corrections in the fiducial cross section \cite{Li:2024ujf}.
In this work, we revisited this issue in $HH \to b \bar{b} \gamma \gamma$, and demonstrated that such large corrections are an artifact of FO calculations which are very sensitive to the interplay of the extra gluon radiation with the phase space restrictions given by the fiducial cuts. Once parton showers (PS) are included, the infrared sensitivity is cured, since the large logarithms arising from gluon radiation are effectively resummed.

We began by showing that the $\gamma\gamma$ system is trivial with respect to NLO QCD corrections. The large FO NLO corrections to the fiducial cross sections are thus entirely due to the $H\to b \bar b$ decay, and would equally arise if $\gamma\gamma$ were replaced by any other final state without colored particles.
We then quantified the effect of NLO QCD corrections to the fiducial cross section, both at FO and with showers. At FO, we found a correction of $15.1\%$ in modulus. Crucially, once PS are included, this sizable FO correction is reduced to only 
$2.9\%$, showing that the large NLO effects in the fiducial cross section are largely washed out by showering.
Furthermore, we established that the same behavior holds at the level of differential cross sections; we illustrated this explicitly using the transverse momentum of the leading $b$-jet and the invariant mass of the $b\bar b \gamma\gamma$ system, both of which exhibit small corrections and stable shapes once PS are taken into account.

To better understand the origin of the large FO corrections, we analyzed the dependence on two fiducial parameters: the minimum jet transverse momentum $p_T^{\min}$ and the jet radius $R$. As expected, large $p_T^{\min}$ values and small $R$ lead to stronger FO corrections, since this severely restricts the phase space available for soft gluon radiation.
In contrast, once PS are included, the NLO corrections become small and largely insensitive to either $p_T^{\min}$ or $R$.

Several directions for future work naturally follow from this study. First, a complete NLO QCD+PS analysis of $gg \to HH \to b \bar b \gamma \gamma$ would be of interest, including a detailed assessment of theoretical uncertainties. It would also be valuable to explore the case where the $\gamma\gamma$ system is replaced by colored particles. Finally, an Effective Field Theory analysis of the full production and decay chain could provide further insight, enhancing the possibility of disentangling genuine QCD effects from possible contributions of new physics.

%% file: acknowledgments.tex
We thank Stefan Gieseke, Stefan Kiebacher and Simon Plätzer for help with \textsc{Herwig}, Silvia Ferrario-Ravasio for help with \textsc{Powheg} and Jian Wang for clarifications on ref.~\cite{Li:2024ujf}.
D.F. is also grateful to Melissa van Beekveld, Kirill Melnikov, Rudi Rahn, Raoul Röntsch, Hua-Sheng Shao and Robert Szafron for discussions. G.H. would like to thank Stephen Jones, Matthias Kerner and Ludovic Scyboz for continuous collaboration on the {\tt ggHH} code. This work was supported by the German Federal Ministry of Research, Technology and Space (BMFTR) under project 05H24VKB, as well as by the Deutsche Forschungsgemeinschaft (DFG, German Research Foundation) under grant 396021762 - TRR 257. The authors also acknowledge support by the state of Baden-Württemberg through bwHPC.